\documentclass{pasj00}

\begin{document}
\SetRunningHead{Sawada et al.}{OTF Observing System of
  NRO 45-m and ASTE 10-m Telescopes}
\Received{2007/10/22}
\Accepted{2007/11/30}

\title{On-The-Fly Observing System of the Nobeyama 45-m and
ASTE 10-m Telescopes}

\author{Tsuyoshi \textsc{Sawada}\altaffilmark{1}, %
  Norio \textsc{Ikeda}\altaffilmark{1,2,3},
  Kazuyoshi \textsc{Sunada}\altaffilmark{1,4},
  Nario \textsc{Kuno}\altaffilmark{1},
  Takeshi \textsc{Kamazaki}\altaffilmark{5},
  Koh-Ichiro \textsc{Morita}\altaffilmark{1,5},
  Yasutaka \textsc{Kurono}\altaffilmark{1,6},
  Norikazu \textsc{Koura}\altaffilmark{7},
  Katsumi \textsc{Abe}\altaffilmark{7},
  Sachiko \textsc{Kawase}\altaffilmark{7},
  Jun \textsc{Maekawa}\altaffilmark{8},
  Osamu \textsc{Horigome}\altaffilmark{9},
  and
  Kiyohiko \textsc{Yanagisawa}\altaffilmark{9}
}

\altaffiltext{1}{Nobeyama Radio Observatory,
  National Astronomical Observatory,\\
  462-2 Nobeyama, Minamimaki, Minamisaku, Nagano 384-1305}
\email{sawada@nro.nao.ac.jp}
\altaffiltext{2}{Institute of Space and Astronautical Science,
  Japan Aerospace Exploration Agency,\\
  Yoshinodai 3-1-1, Sagamihara, Kanagawa 229-8510}
\altaffiltext{3}{Department of Astronomical Science,
  Graduate University for Advanced Studies,
  2-21-1 Osawa, Mitaka, Tokyo 181-8588}
\altaffiltext{4}{Mizusawa VERA Observatory,
  National Astronomical Observatory, 
  2-12 Hoshigaoka, Mizusawa, Oshu, Iwate 023-0861}
\altaffiltext{5}{ALMA-J Project Office,
  National Astronomical Observatory,
  2-21-1 Osawa, Mitaka, Tokyo 181-8588}
\altaffiltext{6}{Department of Astronomy,
  Graduate School of Science,
  The University of Tokyo,
  7-3-1 Hongo, Bunkyo, Tokyo 113-0033}
\altaffiltext{7}{Fujitsu Ltd.,
  1-9-3 Nakase, Mihama, Chiba, Chiba 261-8588}
\altaffiltext{8}{Maekawa Co., Ltd.,
  8240-6078 Nishi-Ide, Oizumi, Hokuto, Yamanashi 409-1501}
\altaffiltext{9}{Fujitsu Nagano Systems Engineering Ltd.,
  1415 Tsurugamidoricho, Nagano, Nagano 380-0813}


%


\maketitle

\begin{abstract}
We have developed spectral line On-The-Fly (OTF) observing mode
for the Nobeyama Radio Observatory 45-m and
the Atacama Submillimeter Telescope Experiment 10-m telescopes.
Sets of digital autocorrelation spectrometers are available
for OTF with heterodyne receivers
mounted on the telescopes, including the focal-plane
$5\times 5$ array receiver, BEARS, on the 45-m.
During OTF observations, the antenna is continuously driven
to cover the mapped region rapidly,
resulting in high observing efficiency and accuracy.
Pointing of the antenna and readouts from the spectrometer
are recorded as fast as 0.1 second.
In this paper we report 
improvements made on software and instruments,
requirements and optimization of observing parameters,
data reduction process, and verification of the system.
It is confirmed that, using optimal parameters,
the OTF is about twice as efficient as conventional
position-switch observing method.
\end{abstract}

\section{Introduction}

On-The-Fly (OTF) is a technique to perform mapping observations
efficiently with single-dish radio telescopes.
In OTF, the antenna is driven continuously in a region to be mapped
and the data are taken in a short interval, instead of
integrating at discrete positions on the sky (such a conventional
``step-and-integrate'' method is hereafter referred to
as {\it position-switch\/} [PSW] observation).
OTF observations have advantages over PSW modes as follows:
(1) Observing efficiency improves since
    dead time of the telescope is reduced and a large number
    ($\gtrsim 10^2$) of on-source integrations are made per
    an emission-free reference integration;
(2) The system variation (e.g., atmospheric conditions,
    receiver gain, pointing of the telescope) less affects the map
    since the entire map can be covered in a short period; and
(3) Since the data are acquired more frequently than
    the Nyquist sampling rate, the spatial information is not lost.
Comprehensive review and discussion on OTF technique have been made
by \citet{mangum2007}.

OTF technique has been commonly and widely used in
radio continuum observations.
Technical limitations, such as fast readout from the spectrometer
and dealing with huge data, had prevented OTF from being applied to
spectral line observations.
In the last decade spectral line OTF has been made practicable
in some observatories
(e.g., \cite{mangum2000,muders2000,ungerechts2000})
and is established as an efficient method.
Likewise, with the Nobeyama Radio Observatory (NRO) 45-m and
Atacama Submillimeter Telescope Experiment (ASTE\footnote{The
ASTE project is driven by NRO,
a branch of National Astronomical Observatory of Japan,
in collaboration with University of Chile,
and Japanese institutes including University of Tokyo,
Nagoya University, Osaka Prefecture University, Ibaraki University,
and Kobe University.}: \cite{ezawa2004,kohno2005}) 10-m telescopes,
continuum OTF observations have already been available.
We enabled the telescopes to be operated in OTF observing mode,
as reported in this paper.
The OTF observing system of the telescopes now works with
an array of digital autocorrelation spectrometers
(MAC; \cite{sorai2000}).
Heterodyne receivers equipped with the telescopes,
including the 25-beam focal plane array receiver
(BEARS: \cite{sunada2000,yamaguchi2000}) on the 45-m telescope,
can be connected to the MAC and be available for OTF observations.

The capability of OTF to obtain a high-quality map efficiently
is important in various scientific fields.
For example, investigating
how the initial mass function (IMF) of stars is determined
is one of the most important issues in astronomy.
Recent studies have suggested that the stellar IMF is
well related to the mass function (MF) of dense molecular cloud cores
(e.g., \cite{ikeda2007}), which are sites of star formation.
Therefore detecting a number of cores in various star forming regions
is needed in order to study the relation between star forming activity
and MF or physical/dynamical conditions of the cores.
In this case sensitive unbiased mapping observations of
wide fields are required.
In addition, OTF data are ideal to be combined with
interferometric data, since they preserve spatial information
in the (spatially) low frequency regime,
which is lost via interferometric observations.
Thus, the fidelity of high resolution images of, e.g.,
molecular cloud cores or external galaxies, taken with interferometers
are significantly improved by being combined with OTF data.
The fidelity (preserving total flux) is essential to various kinds of
studies.
For instance, in order to investigate the evolution of the
interstellar medium (formation and destruction of
giant molecular cloud associations [GMAs])
across spiral arms in external galaxies, it is required
to detect not only GMAs, which are spatially confined to spiral arms,
but also diffuse emission in interarm regions.

In order to make spectral line OTF observations available
with the NRO 45-m and the ASTE 10-m telescopes,
various improvements have been applied to
the control software system, COSMOS-3 \citep{morita2003,kamazaki2005},
and related instruments.
In particular, fast and synchronized control of instruments and
handling of huge amount of data are the difficulties,
as discussed later.
The way to determine optimal observing parameters is also complicated.

In this paper we report practical information and implementation
of the OTF observing mode for the telescopes.
Basic concepts and parameters of OTF observations are introduced
in \S \ref{sec:overview}, followed by a brief description on
controls of related instruments in \S \ref{sec:control}.
Requirements on observing parameters,
procedures to estimate sensitivity, and how to maximize observing
efficiency are explained in \S \ref{sec:parameters}.
We describe the data reduction process in \S \ref{sec:reduction}.
Desired characteristics of convolution functions to regrid the
data onto a map are discussed
and some appropriate functions are shown.
In \S \ref{sec:verification} we describe
the measurements of driving performance of the antennas,
comparison with PSW map, and
verification of frequency-switch observing mode.

\section{Basic Observing Process and Parameters}\label{sec:overview}

\begin{figure}
  \begin{center}
    \FigureFile(80mm,45mm){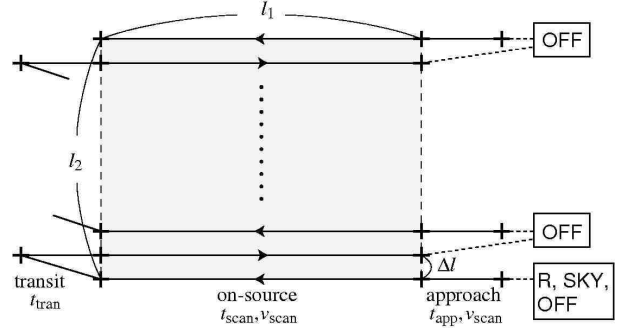}
  \end{center}
  \caption{An example of scan pattern of OTF observations.
    The gray rectangle is a region to be mapped
    with a set of horizontal scans.
    The case of $N_{\rm scan}^{\rm SEQ} = 2$
    (an OFF is inserted before every 2 ON scans)
    is shown.
  }\label{fig:scanpat}
\end{figure}

An OTF scan pattern is schematically illustrated in
Fig.\ \ref{fig:scanpat}.
A constant-speed raster scan with a single-beam receiver is assumed.
Like usual PSW observations,
the standard ``chopper-wheel'' technique \citep{penzias1973,ulich1976}
is employed to calibrate the antenna temperature $T_{\rm A}^*$
in which the atmospheric and antenna losses are corrected.
For the chopper-wheel calibration,
a hot load (R) at the ambient temperature and
blank sky (SKY) are observed during the observation
in an appropriate interval.
Data at an emission-free reference position (OFF) are taken
before every on-source scan (ON, hereafter simply ``scan'')
or set of several scans.
The OFF is usually taken by pointing the antenna outside the source.
However it can also be taken by shifting observing {\it frequency\/}
(see \S \ref{subsubsec:freqsw}).

During the scan the antenna is driven at a constant speed
($v_{\rm scan}$) on the sky and the data are ``dumped'' from the
spectrometer at an interval of $t_{\rm dump}$.
The ``approach-run'' at a speed $v_{\rm scan}$ is inserted
before every scan to let the antenna move stably during the scan.
If there is no OFF between two scans,
a ``transit-run'' (from the end point of the scan
to the start point of the next approach-run) is inserted.

We set fundamental parameters as follows (see Fig.\ \ref{fig:scanpat}).
In \S \ref{sec:parameters} we discuss how to determine them.
The dimension of the mapping area is $l_1\times l_2$
($l_1$ along the scan, $l_2$ across the scan).
The scan speed of the antenna on the sky is $v_{\rm scan}$.
Time to be taken in an approach-run,
a main (on-source) scan, and a transit-run are
$t_{\rm app}$, $t_{\rm scan} = l_1/v_{\rm scan}$, and
$t_{\rm tran}$, respectively.
A separation between the scan rows is ${\mit\Delta}l$,
the number of scans taken between a pair of OFFs is
$N_{\rm scan}^{\rm SEQ}$,
and the grid spacing of a map to be made is $d$.
Time to be taken to slew the antenna between the mapped region
and the OFF position is $t_{\rm tran}^{\rm OFF}$.
The parameters $t_{\rm app}$, $t_{\rm tran}$, and
$t_{\rm tran}^{\rm OFF}$ depend on the performance of the antenna,
the scan speed $v_{\rm scan}$, position on the sky, etc.,
and are typically in the range of a few to 10 seconds.
Measurements of the performance of the antenna are described in
\S \ref{subsec:antperformance}.
The time interval between each dump from the spectrometer is
$t_{\rm dump}$.

\section{Control of Related Instruments}\label{sec:control}

In OTF, spectral data are dumped at short interval ($\sim 0.1$ second)
while the antenna is continuously driven across the sky.
Consequently the following difficulties arise.
\begin{description}
  \item[Antenna.] The antenna should be smoothly driven on
    a scan path. The sky coordinates, toward which
    the antenna pointed, have to be recorded for each data dump.
  \item[Spectrometer.] Data dump is done at short interval,
    and it must be synchronized with the antenna driving.
  \item[Local oscillator.] Doppler correction of the telescope
    with respect to a rest frame and frequency switch (FSW)
    are difficult to be implemented in a conventional way,
    since exceedingly rapid control of the frequency is required.
  \item[Data production and storage.] High data production rate,
    caused by the fact that the data are dumped $\sim 10^2$
    times more frequently than typical PSW observations,
    may cause trouble on data storage and reduction process.
\end{description}

In this section we describe the control system and instruments
to resolve these issues.

\subsection{COSMOS-3}

\begin{figure}
  \begin{center}
    \FigureFile(80mm,59mm){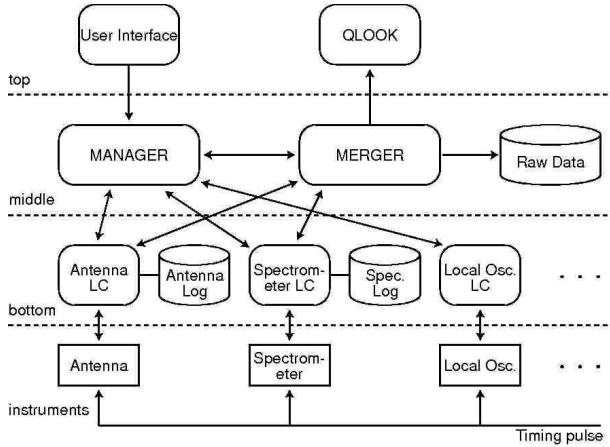}
  \end{center}
  \caption{Schematic illustration of three-layered COSMOS-3
    architecture.
    Rounded rectangles are components of COSMOS-3,
    cylinder symbols are data storage and files therein, and
    rectangles are individual instruments.
  }\label{fig:cosmos3}
\end{figure}

The control software system of the telescopes is
COSMOS-3 \citep{morita2003}.
It is designed to have a three-layer architecture,
as illustrated in Fig.\ \ref{fig:cosmos3}.
The interface with observers is provided in the top layer.
In the middle layer, MANAGER and MERGER exist.
The MANAGER controls the observation by
receiving instruction from observers and interacting with
Local Controllers (LCs), which directly control instruments,
in the bottom layer.
The MERGER receives outputs from LCs and produces calibrated data.
The observer can view a Quick Look (QLOOK) of the data
and check if the observation is running properly.
Synchronization of instruments with each other is managed
with 1, 10, 50, and 100-Hz timing pulses,
which are generated and distributed from the standard clock
system of the observatory.

\subsection{Antenna}

Before every approach-run, the antenna LC instructs the antenna
to settle on the starting point of the approach-run.
When the antenna is judged to be steady (the positional error
has been within a tolerance for a specified duration),
the MANAGER decides the timing and coordinates of the following
approach-run(s), scan(s), and transit-run(s).
The antenna LC instructs the antenna to follow an ideal scan path
at the decided timing.
The ideal coordinates, toward which the antenna is desired to point,
is hereafter called PROG values.
The PROG values for every 0.1 second are calculated and transmitted
to the antenna, 0.2 second before it is used.
The antenna manages the timing to use the transmitted coordinates
with the 10-Hz (0.1-second) timing signal.
Two neighboring PROG values are linearly interpolated.
The antenna is driven to follow the interpolated positions,
referring the 50 and 100-Hz timing signals.

Meanwhile, the actual pointing of the antenna (hereafter
referred to as REAL values) jitters around the PROG values.
The amount of the jitter (${\rm PROG}-{\rm REAL}$) is measured,
in case of the 45-m telescope, using the encoder readout of
the master collimator and the pointing deviation between
the master collimator and the main reflector.
In case of the 10-m telescope, the encoder readout of the antenna
is used.
The PROG (in both equatorial and horizontal coordinates) and
${\rm PROG}-{\rm REAL}$ (horizontal coordinates) values are
both written onto a file, the {\it Antenna Log\/} file,
along with time stamps at every 0.1 second.

\subsection{Spectrometer}

It is essential to synchronize the timing of data acquisition
with that of the antenna driving,
particularly in case of OTF, since the data dump time is very short.
The timing is managed as follows.

When an observation starts, observing parameters,
such as interval of data dumps (0.1 second) and
number of datasets to be dumped, are calculated by the MANAGER
from instructions made by the observer.
They are sent to the spectrometer LC and then to the spectrometer,
in advance of data acquisition.
For each scan, the MANAGER decides the time
to start the scan as mentioned above.
The LC is informed of the time, waits until 1 second before it,
and then sends a command to the spectrometer to start integration.
After receiving it, the spectrometer begins to acquire the data,
triggered by the next fall-down of the 1-Hz, 50\%-duty timing pulse.
It dumps the data at every 0.1 second during the scan.

Though the spectrometer outputs datasets at every 0.1 second,
observers can specify $t_{\rm dump}$ to be $0.1N$ ($N=1,2,...$)
seconds to reduce the data size.
Every 0.1-second output is transferred to the LC.
The LC averages successive $N$ datasets if $N>1$, and
writes them down onto a file (the {\it Spectrometer Log\/} file)
along with time stamps.

\subsection{Local Oscillator}

\subsubsection{Doppler Tracking}

When a wide area is mapped, the radial velocity
of the telescope ($v_{\rm rad}$) with respect to
the reference frame (e.g., the Local Standard of Rest [LSR])
significantly changes from point to point:
the velocity gradient across the sky plane amounts to roughly
$0.8\;{\rm km\; s^{-1}\; deg^{-1}}$ (LSR) at maximum in NRO.
In PSW observations,
the radial velocity of the telescope at each position is
calculated and the local oscillator (LO) frequency is shifted
before the integration starts, in order to track
the Doppler shift caused by ${\mit\Delta}v_{\rm rad}$.

However, in OTF observations, the antenna runs on a scan path
(thus $v_{\rm rad}$ gradually changes),
and during which the spectral data are continuously taken
at a short interval.
In this case it is difficult to track $v_{\rm rad}$
by shifting the LO frequency continuously during the scan.
Therefore we do not make any shift on LO frequency
during the OTF scans.
Instead, the Doppler correction is carried out by software:
in the production process of spectral data (MERGER; see below),
$v_{\rm rad}$ for each ON integration is calculated,
and then channel shift operation is made.

\subsubsection{Frequency Switching}\label{subsubsec:freqsw}

When spatially-widespread and/or narrow lines
(e.g., CO in nearby molecular clouds) are mapped,
FSW observing technique has been widely used.
Conventional FSW observations with the NRO 45-m and
ASTE 10-m telescopes are performed as follows.
The LO frequency is switched between
the original value $\nu_{\rm LO}^0$ and slightly shifted value
$\nu_{\rm LO}^0 \pm {\mit\Delta}\nu_{\rm LO}$.
The switching cycle is in the order of $\sim 1$ Hz, and
the frequency throw ${\mit\Delta}\nu_{\rm LO}$ is typically
several MHz.
The averager/integrator of the spectrometer behaves in
adding and subtracting mode, synchronized with the frequency shift
of the LO, during the ON integration
(the averager/integrator does nothing during a margin for
the transition of frequency).
Thus the ON concurrently plays a role of OFF.
Accordingly the original signal and
a negative-amplitude signal appear on the spectrum
at a separation of ${\mit\Delta}\nu_{\rm LO}$.
The final spectrum is demodulated by invert, shift and add operation.

In OTF, the difficulty of shifting LO frequency
during a short integration again arises.
We have implemented FSW observing mode in OTF
by shifting the LO frequency at OFF position only:
$\nu_{\rm LO}^0$ is used during ON,
while $\nu_{\rm LO}^0 \pm {\mit\Delta}\nu_{\rm LO}$ is used during OFF.
The emission line existing (if any) at OFF position shifts
by ${\mit\Delta}\nu_{\rm LO}$,
thus it can be virtually regarded as emission-free reference.

\subsection{Spectral Data Production}

As described above, the antenna pointings are recorded
in the Antenna Log file at every 0.1 second, and
the readouts from the spectrometer are written onto
the Spectrometer Log file at every $0.1N$ seconds.
The MERGER, which is a process to produce spectral data files,
runs during the observation
(it can also be executed asynchronously after the observation).
The MERGER process reads the Spectrometer Log file,
calculates the standard chopper-wheel equation to
calibrate the antenna temperature,
and shifts the resultant spectra along the frequency channels
to correct Doppler shift.
The MERGER also reads the Antenna Log file.
By comparing time stamps written in both log files,
it calculates sky coordinates, toward which the antenna
was actually pointed at the moment of each dump,
using the PROG (equatorial) and ${\rm PROG}-{\rm REAL}$
(horizontal) values.
The derived coordinates are written in the header of
the spectral data file.
In this way intensity- and velocity-calibrated spectra
are written on a file, called the {\it Raw Data\/} file.

In practical implementation, the trouble with the Raw Data
is their large sizes.
If the 25-beam, 1024-channel data are represented as 32-bit
and dumped every 0.1 second, the data rate goes up to
$\simeq 3.5\;{\rm GB\;hour^{-1}}$ (data headers are neglected).
In order to reduce the size of the Raw Data,
the quantization level, $Q$ [bit], must be as small as possible.
When the peak antenna temperature of the intrinsic spectrum
is $T_{\rm p}$ and rms noise is $\sigma$,
observed antenna temperature between
the minimum ($-N\sigma$; $N\simeq 3$) and maximum
($T_{\rm p}+N\sigma$) is quantized into $2^Q$ levels.
The quantization noise is roughly written as
$\sigma_Q = (T_{\rm p}+2N\sigma)/2^{Q+1}$.
Since the noise level increases to $\sqrt{\sigma^2+\sigma_Q^2}$,
the overhead in the observing time
(the ratio between $Q$-bit and $\infty$-bit cases) is
\begin{equation}
  \frac{t_Q}{t_\infty} = 1 +
    \left( \frac{T_{\rm p}+2N\sigma}{2^{Q+1}\sigma} \right)^2 ,
\end{equation}
meaning that the loss increases when the signal-to-noise ratio
of individual spectrum, $T_{\rm p}/\sigma$, is high.
Supposing an extreme case for the current system:
peak antenna temperature $T_{\rm p} = 100\;[{\rm K}]$,
system noise temperature $T_{\rm sys} = 100\;[{\rm K}]$,
frequency resolution $B = 1\;[{\rm MHz}]$,
and data dump time $t_{\rm dump} = 0.1\;[{\rm s}]$,
we obtain $t_Q/t_\infty-1 \simeq (320/2^{Q+1})^2$.
For $Q=8$, 12, and 16, $t_Q/t_\infty-1$ becomes
0.40, $1.5\times 10^{-3}$, and $6.1\times 10^{-6}$,
respectively.
We adopted $Q=12$ (4096 levels), for which
the quantization noise is practically negligible,
resulting in the maximum data rate of
$\simeq 1.3\;{\rm GB\;hour^{-1}}$.

In addition, the MERGER supports options of
channel trimming and channel binning.
When these options are specified by the observer,
the MERGER trims and/or bins up channels of the spectra
before writing them down on the Raw Data file.

It should be noted that the 12-bit quantization may affect
the data in the following situation:
bandpass becomes nearly 0 (i.e., $T_{\rm A}^*$ diverges)
at the band edges,
or an extremely intense spurious signal appears.
The observer should pay attention to the obtained spectra
through QLOOK.
This problem can be evaded by trimming the band edges
and/or spurious signals, since the data are quantized
{\it after\/} the trim.

\section{Observing Parameters}\label{sec:parameters}

\subsection{Requirements on Sampling}

As a result of an observation, the mapped region is filled with
data points.
The data sampling separation is $v_{\rm scan} t_{\rm dump}$ along
the scan, ${\mit\Delta}l$ across the scan.
\citet{mangum2007} discussed requirements on the sampling,
as summarized as follows.
It is requested that, at least, the sampling rates
$v_{\rm scan} t_{\rm dump}$ and ${\mit\Delta}l$ are both
more frequent than the Nyquist sampling rate $\lambda/2D$,
where $\lambda$ is the observed wavelength and $D$ is the
diameter of the antenna aperture.
In case of $\lambda=2.6\;{\rm mm}$ (115 GHz) observations with
the 45-m telescope $\lambda/2D \simeq 6\arcsec$, which corresponds to
$\simeq 1/2.5$ of the half-power beam width (HPBW).
Practically a sampling more frequent than the Nyquist rate
is required to avoid aliasing noise and beam smearing effects.
Since the data points do not align on any regular grid
due to the antenna jitter etc.,
the data should be regridded onto a regular grid
using a gridding convolution function (GCF) in
the data reduction process (see \S \ref{sec:reduction}).

\subsection{Estimation of Sensitivity}

We estimate the sensitivity of a single-beam observation.
Application to a multi-beam receiver is discussed later.

The total number of scan rows in an observation is
$N_{\rm row} = l_2/{\mit\Delta}l + 1$.
The total on-source integration time becomes
\begin{equation}
  t_{\rm tot}^{\rm ON} = N_{\rm row} t_{\rm scan} .
\end{equation}
The total time spent to run an observation including
R, SKY, OFF, antenna slew, etc.\ is estimated to be
\begin{equation}
  t_{\rm tot}^{\rm OBS} = N_{\rm row}
    \left( t_{\rm scan} + t_{\rm OH} +
    \frac{t_{\rm OFF}}{N_{\rm scan}^{\rm SEQ}}
    \right) f_{\rm cal} ,
\end{equation}
where $t_{\rm OFF}$ is an integration time for an OFF,
and $f_{\rm cal}$ is an overhead of R-SKY calibration
(if 1 minute is consumed to obtain R and SKY data
at every 15 minutes, $f_{\rm cal} = 16/15$).
The $t_{\rm OH}$ is an overhead time per one scan row,
which consists of go-and-return to the OFF point
$2t_{\rm tran}^{\rm OFF}$,
time for approach- and transit-run
$t_{\rm app}$ and $t_{\rm tran}$, thus is written as
\begin{equation}
  t_{\rm OH} = \frac{2t_{\rm tran}^{\rm OFF}}{N_{\rm scan}^{\rm SEQ}}
    + t_{\rm app} +
    \frac{N_{\rm scan}^{\rm SEQ}-1}{N_{\rm scan}^{\rm SEQ}}
    t_{\rm tran} .
\end{equation}
The $t_{\rm tran}^{\rm OFF}$ depends on the distance
between the mapped region and the OFF position;
$t_{\rm app}$ and $t_{\rm tran}$ should be chosen so that
the antenna follows the PROG position during the scans.
Now the ratio of on-source time to the total time spent is
\begin{eqnarray}
  \eta_{\rm ON/OBS} &=& \frac{t_{\rm tot}^{\rm ON}}
      {t_{\rm tot}^{\rm OBS}}\\
    &=& \frac{t_{\rm scan}}
      {t_{\rm scan}+t_{\rm OH}+t_{\rm OFF}/N_{\rm scan}^{\rm SEQ}}
    \cdot \frac{1}{f_{\rm cal}} .
\end{eqnarray}

The total on-source integration time for a map grid point,
$t_{\rm cell}^{\rm ON}$, is a sum of time during which
the beam scans within the grid cell.
Since the data are convolved using a GCF to construct
a regularly gridded map,
effectively a factor $\eta$ is multiplied:
\begin{eqnarray}
  t_{\rm cell}^{\rm ON} &=&
    \eta \cdot \frac{d^2}{l_1 l_2} t_{\rm tot}^{\rm ON} \\
    &\simeq& \frac{\eta t_{\rm scan}d^2}{l_1{\mit\Delta}l} .
\end{eqnarray}
The factor $\eta$ is a constant determined by the extent of
the used GCF and is calculated as follows.
Suppose that observed points $i=1,2,...$ are uniformly distributed
around the grid point and each point has a spectrum $T_i(k)$
($k=1,...,N_{\rm CH}$), rms noise temperature $\sigma_i$,
and a GCF weight $w_i$.
We assume that the on-source integration time $t_0$
and therefore the rms noise temperature
$\sigma_i = \sigma_0 = T_{\rm sys}/(\eta_{\rm q} \sqrt{B\, t_0})$
of each point are both constant.
Here $T_{\rm sys}$ is the system noise temperature,
$B$ is the frequency resolution of the spectra,
and $\eta_{\rm q}$ is the quantization efficiency of the spectrometer.
In case of MAC $\eta_{\rm q}=0.88$; and hereafter $\eta_{\rm q}$ is
omitted from expressions.
The convolved spectrum $T(k)$ is written as
$T = (\sum w_i T_i)/(\sum w_i)$,
and its noise temperature $\sigma$ becomes
$\sigma = (\sqrt{\sum w_i^2}/\sum w_i) \sigma_0
  = T_{\rm sys}/\sqrt{B t_{\rm cell}^{\rm ON}}$,
where $t_{\rm cell}^{\rm ON} \equiv t_0\, (\sum w_i)^2/\sum (w_i^2)$.
If we take the grid spacing as the unit of spatial length and
re\-define $t_0$ as the on-source integration time per unit area
(1 grid cell), summations can be rewritten with integrals:
$t_{\rm cell}^{\rm ON} = t_0\, (\int w\, dx dy)^2/\int w^2\, dx dy
  \equiv \eta t_0$.
Approximate values of $\eta$ for GCFs Bessel$\times$Gauss,
Sinc$\times$Gauss, Gauss, Pillbox, and Spheroidal
(see \S \ref{sec:reduction}) with default parameters are, respectively,
4.3, 1.2, 6.3, 1.0, and 10.

Redefining $B$ as the frequency resolution of a {\it map\/} to be made,
the noise of the map due to on-source integration is estimated to be
\begin{equation}
  {\mit\Delta}T_{\rm A}^*({\rm ON}) =
    \frac{T_{\rm sys}}{\sqrt{B\, t_{\rm cell}^{\rm ON}}} ,
\end{equation}
the standard radiometer equation.
On the other hand, the number of OFF points used to consist
a map grid point is roughly written as $d/{\mit\Delta}l$
(here the extent of the GCF is neglected).
Thus the effective OFF integration time for a grid cell is
\begin{equation}
  t_{\rm cell}^{\rm OFF} \simeq \frac{d}{{\mit\Delta}l} t_{\rm OFF} ,
\end{equation}
and the noise due to OFF points becomes
\begin{equation}
  {\mit\Delta}T_{\rm A}^*({\rm OFF}) =
    \frac{T_{\rm sys}}{\sqrt{B\, t_{\rm cell}^{\rm OFF}}} .
\end{equation}
Therefore the total noise level of the map is written as
\begin{eqnarray}
  {\mit\Delta}T_{\rm A}^* &=&
    \sqrt{{\mit\Delta}T_{\rm A}^*({\rm ON})^2 +
      {\mit\Delta}T_{\rm A}^*({\rm OFF})^2}\\
    &=& \frac{T_{\rm sys}}{\sqrt{B}}
    \sqrt{\frac{1}{t_{\rm cell}^{\rm ON}} +
      \frac{1}{t_{\rm cell}^{\rm OFF}}} .
    \label{eq:dTa}
\end{eqnarray}
Validity of this equation has been practically confirmed
(see \S \ref{subsec:comparepsw}).

\subsection{Optimization}

The noise level of a map achieved in unit observing time
${\mit\Delta}T_{\rm A}^*(0)$ is written as
\begin{eqnarray}
  {\mit\Delta}T_{\rm A}^*(0) &=&
    {\mit\Delta}T_{\rm A}^* \sqrt{t_{\rm tot}^{\rm OBS}}
    \label{eq:dTa0}\\
    &=&
    \frac{T_{\rm sys}}{\sqrt{B}}
    \sqrt{\left( \frac{1}{t_{\rm cell}^{\rm ON}} +
    \frac{1}{t_{\rm cell}^{\rm OFF}} \right)} \times \nonumber \\
    &&
    \sqrt{\left( t_{\rm scan} + t_{\rm OH} +
    \frac{t_{\rm OFF}}{N_{\rm scan}^{\rm SEQ}} \right)
    N_{\rm row} f_{\rm cal}}
\end{eqnarray}
and is minimized when $t_{\rm OFF}$ is optimal:
\begin{equation}
  \frac{\partial}{\partial t_{\rm OFF}} {\mit\Delta}T_{\rm A}^*(0) = 0
  \label{eq:ddTa0dt}
\end{equation}
leads to
\begin{equation}
  t_{\rm OFF}^{\rm optimal} \simeq
    \sqrt{\left( t_{\rm scan}+t_{\rm OH} \right)
    \frac{\eta d\, t_{\rm scan}}{l_1}}
    \sqrt{N_{\rm scan}^{\rm SEQ}} .\label{eq:toffopt}
\end{equation}
This formula is a generalization of the well-known relation
$t_{\rm OFF}=\sqrt{N} t_{\rm ON}$
for PSW observations,
where $N$ is the number of ONs taken per one OFF.
Toward $t_{\rm OH}\to 0$, Eq.\ (\ref{eq:toffopt})
resolves itself into the $\sqrt{N}$-relation.

\begin{figure}
  \begin{center}
    \FigureFile(80mm,58mm){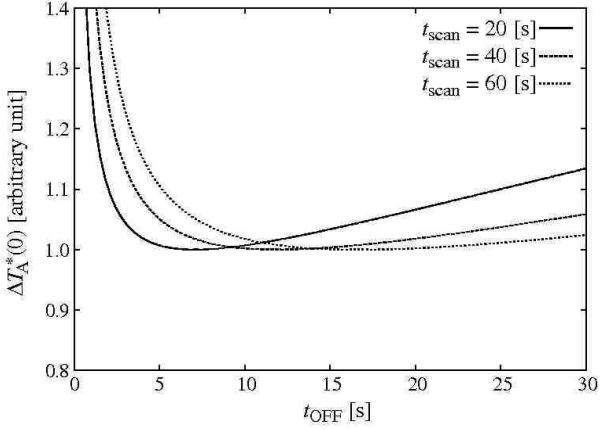}
  \end{center}
  \caption{Dependence of ${\mit\Delta}T_{\rm A}^*(0)$
    (normalized by the minimum)
    on $t_{\rm OFF}$ are shown in three cases:
    $t_{\rm scan}=20$ ({\it solid\/}), 40 ({\it dashed\/}),
    and 60 ({\it dotted\/}) seconds.
    Common parameters are: $l_1=600\;[\arcsec]$,
    ${\mit\Delta}l=5\;[\arcsec]$, $d=7.5\;[\arcsec]$,
    $t_{\rm OH}=25\;[{\rm s}]$, and $N_{\rm scan}^{\rm SEQ}=1$.
    Bessel$\times$Gauss convolution ($\eta=4.3$) is assumed.
  }\label{fig:toff}
\end{figure}

Dependence of ${\mit\Delta}T_{\rm A}^*(0)$ on $t_{\rm OFF}$ is
plotted in Fig.\ \ref{fig:toff}.
For cases of $t_{\rm scan}=20$, 40, and 60 seconds
(other parameters are shown in the caption),
${\mit\Delta}T_{\rm A}^*(0)$ is minimized at
$t_{\rm OFF}^{\rm optimal}=7$, 12, and 17 seconds, respectively.
If $t_{\rm OFF}$ is shorter than the optimal value,
${\mit\Delta}T_{\rm A}^*({\rm OFF})$ dominates the map.
On the other hand, $t_{\rm OFF}$ longer than
$t_{\rm OFF}^{\rm optimal}$ is excessive since the noise level
of the map is limited by ${\mit\Delta}T_{\rm A}^*({\rm ON})$.

Using the above notations, it is quantitatively shown how
the observing efficiency improves, by adopting appropriate parameters,
compared with PSW observations.
The following two factors contribute:
%
(1) The ratio of on-source time to the total time spent,
    $\eta_{\rm ON/OBS}$, becomes larger because
    not only the dead time (antenna slew, etc.)\ is reduced,
    but also the OFF integration time is relatively shorter
    ($t_{\rm OFF} \ll t_{\rm scan}$); and
(2) In general $t_{\rm cell}^{\rm ON} \ll t_{\rm cell}^{\rm OFF}$,
    thus ${\mit\Delta}T_{\rm A}^*$ nearly equals to
    ${\mit\Delta}T_{\rm A}^*({\rm ON})$ (instead of
    ${\mit\Delta}T_{\rm A}^*\simeq
    \sqrt{2}{\mit\Delta}T_{\rm A}^*({\rm ON})$,
    which is applicable to PSW observation
    with $t_{\rm ON}=t_{\rm OFF}$).
%
As $t_{\rm OH}\to 0$ and $t_{\rm scan}\to \infty$,
the both factors respectively correspond to reduction of
observing time by a factor of 2.
Thus OTF is, theoretically, up to 4 times
more efficient than PSW.
In practice, improvement of efficiency amounts to
a factor of $\sim 2$.

\subsection{Application to an Array Receiver}

The above discussion is made for a single-beam receiver.
In case of an array (multi-beam) receiver, some expressions change.
Here we consider the case of BEARS, a $5\times 5$ focal plane array.

\begin{figure}
  \begin{center}
    \FigureFile(80mm,42mm){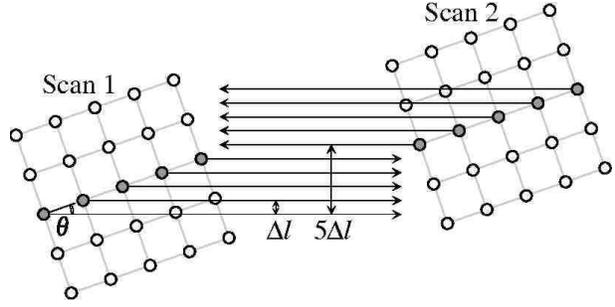}
  \end{center}
  \caption{A schematic illustration of OTF scan patterns with BEARS.
    The array is inclined by an angle $\theta$ with respect to
    the direction of scans in order to make the scans by neighboring
    beams separated by ${\mit\Delta}l$.
    The first scan (labeled as ``Scan 1'') and the next
    (``Scan 2'') are offset by $5{\mit\Delta}l$ (see text).
  }\label{fig:scanpatbears}
\end{figure}

Fig.\ \ref{fig:scanpatbears} schematically shows OTF scans with BEARS
considered here.
The array is inclined with respect to the scan direction by an angle
$\theta$: a neighboring couple of beams makes a pair of scans
separated by a distance ${\mit\Delta}l = L\sin\theta$
($L=\timeform{41."1}$ is a beam separation).
The next scan runs $5{\mit\Delta}l$ away, thus
the number of scan rows is written as
$N_{\rm row} = l_2/(5{\mit\Delta}l)+1$.
In this case scans made by 5 beams in each row of the array fill
the mapped region at a separation of ${\mit\Delta}l$.
Since the 5 rows of the array respectively cover the map,
\begin{equation}
  t_{\rm cell}^{\rm ON} =
    \frac{5\eta t_{\rm scan}d^2}{l_1{\mit\Delta}l}
  \label{eq:tcellONbears}
\end{equation}
and
\begin{equation}
  t_{\rm cell}^{\rm OFF} = \frac{5d}{{\mit\Delta}l}t_{\rm OFF} .
  \label{eq:tcellOFFbears}
\end{equation}
The noise level of the map is obtained by substituting
$t_{\rm cell}^{\rm ON}$ and $t_{\rm cell}^{\rm OFF}$ in
Eq.\ (\ref{eq:dTa}) with Eqs.\ (\ref{eq:tcellONbears}) and
(\ref{eq:tcellOFFbears}).
Following the transformation in the previous subsection, we have
\begin{equation}
  t_{\rm OFF}^{\rm optimal} =
    \sqrt{\left( t_{\rm scan}+t_{\rm OH} \right)
    \frac{\eta d\, t_{\rm scan}}{l_1}}
    \sqrt{N_{\rm scan}^{\rm SEQ}} ,
\end{equation}
the same expression as Eq.\ (\ref{eq:toffopt}).

The strategy mentioned above focuses on mapping speed:
compared with the case of a single-beam receiver,
the mapped region can be covered 5 times faster and
the integration becomes 5 times deeper.
There may be another strategy,
conscious of the uniformity of the map.
Using the same parameters as the case of a single-beam receiver
(each beam runs at a separation of ${\mit\Delta}l$),
the integration simply becomes 25 times deeper
in the same observing time.
The point is that every scan path is approximately traced by
{\it all\/} the 25 beams:
characteristics of the beams are expected to be averaged out.
However whether this advantage is realized or not depends
on the condition, since the system variation may spoil the uniformity
if the observing time becomes too long.

\section{Data Processing}\label{sec:reduction}

\subsection{Data Reduction}

Reduction process of OTF data is done with {\it NOSTAR\/}
(Nobeyama OTF Software Tools for Analysis and Reduction)
developed in NRO.
It is designed to run on UNIX or UNIX-like operating systems.
Its core functions (baseline subtraction,
creating cube FITS from spectra, etc.)\ are
provided as command-line tools written in C/FORTRAN
in order to enable batch processing.
These command-line tools are wrapped in graphical user interface (GUI)
written in the Interactive Data Language (IDL).

A run of observation produces a Raw Data file
(see \S \ref{sec:control}).
Spectra from all the used spectrometers are contained within it.
The first step of data reduction process is to extract the data
to the user's working directory.
The task, named {\it Split\/}, does not simply copy the Raw Data file,
but divide it into {\it Split Raw Data\/} files according to
the spectrometers.
Namely, a Raw Data taken with BEARS is split into
25 Split Raw Data files.
Accordingly the following process can be separately applied
for each Split Raw Data.

Subsequent procedures of data reduction
(baseline subtraction, bad data flagging, etc.)\ are
basically the same as those for PSW data,
except for large size of the data.
The GUI is designed to be batch-oriented,
to help users to proceed the reduction process rapidly
without being bothered with a large quantity of data.
Each process overwrites the Split Raw Data itself
in order to avoid running out of space of the working directory.

Finally a map (cube FITS) is made from the processed spectra.
An OTF observation (or a series of observations) produces
a set of data points which fills the mapped region
with spacings smaller than the Nyquist sampling rate,
as described in \S \ref{sec:parameters}.
The data are convolved into the map using a GCF.
Desired characteristics of GCFs and appropriate function forms
are discussed below.
The obtained map may suffer from the so-called {\it scanning noise\/}
along the scan direction, in addition to the statistical noise.
The scanning noise can be effectively removed by combining
two maps made from orthogonal scans using the so-called
{\it basket-weave\/} method.
We have implemented the
{\it PLAIT\/} algorithm described by \citet{emerson1988}.

\subsection{Gridding Convolution Functions}\label{subsec:convolution}

GCFs used to make maps are desired to have
the following characteristics.

First: form of GCF is similar to that of the telescope beam itself.
Convolution with such a GCF corresponds to,
in the Fourier domain, that a weighting function
(Fourier-transformed beam pattern) is multiplied twice
to the intrinsic spatial frequency distribution.
Consequently the best signal-to-noise ratio is achieved.
Though we cannot know the beam pattern at infinite accuracy in
practice, one should choose a GCF so that it mimics the telescope beam.

Second: the GCF's energy concentration ratio is high.
In practical convolution operation, the extent of the GCF is finite.
Thus, in the Fourier domain, the GCF has artificial frequency
components ({\it sidelobes\/}).
By regridding the data, these sidelobes are folded onto
the primary component, resulting in the so-called
{\it aliasing noise}.
In order to preserve the observed spatial frequency information,
the aliasing effect must be as small as possible.
As an index, energy concentration ratio
\begin{equation}
  {\cal R} = \frac{\int_A |C(\eta)|^2 d\eta}
    {\int_{-\infty}^{\infty} |C(\eta)|^2 d\eta}
\end{equation}
is introduced \citep{briggs1999}, where the GCF is $c(l)$,
$l=x/{\mit\Delta}x$, ${\mit\Delta}x$ is the grid spacing,
$C(\eta)$ is the Fourier transform of $c(l)$,
and $A$ is the area in which $\eta<1$.
Here ${\cal R}$ represents the degree of concentration
of the GCF within $A$: it is expected that the larger ${\cal R}$ is,
the smaller the aliasing effect is.

\begin{figure}
  \begin{center}
    \FigureFile(80mm,59mm){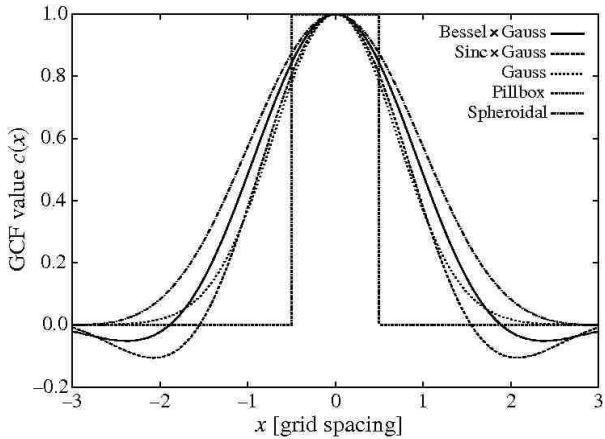}
  \end{center}
  \caption{Shapes of implemented GCFs $c(x)$:
    Bessel$\times$Gauss ({\it solid\/}),
    Sinc$\times$Gauss ({\it dashed\/}),
    Gauss ({\it short-dashed\/}),
    Pillbox ({\it dotted\/}), and
    Spheroidal ({\it dot-dashed\/});
    normalized so that $c(0)=1$.
    Function parameters are their default ones (see text).
  }\label{fig:gcfs}
\end{figure}

The following GCFs (their shapes are shown in Fig.\ \ref{fig:gcfs})
are implemented so far:

\begin{description}

\item[Bessel$\times$Gauss:]
a Gaussian-tapered Jinc function
\begin{equation}
  c(r) = \left\{
    \begin{array}{ll}
      \frac{J_1(\pi r/a)}{\pi r/a}
        \exp \left[ -\left( \frac{r}{b} \right)^2 \right] &
        (r \le R_{\rm max}) \\
      0 & (\mbox{otherwise})
    \end{array}
    \right.
\end{equation}
where $J_1$ is the 1st-order Bessel function,
$r$ is the distance between the data point and the grid point
(the unit is the grid spacing).
Parameters $a$, $b$, and $R_{\rm max}$ can be arbitrary chosen,
and are set as $a=1.55$, $b=2.52$, $R_{\rm max}=3$ by default
(see below).

\item[Sinc$\times$Gauss:]
a Gaussian-tapered Sinc function
\begin{equation}
  c(r) = \left\{
    \begin{array}{ll}
      \frac{\sin(\pi r/a)}{\pi r/a}
        \exp \left[ -\left( \frac{r}{b} \right)^2 \right] &
        (r \le R_{\rm max}) \\
      0 & (\mbox{otherwise})
    \end{array} .
    \right.
\end{equation}
The default values for parameters $a$, $b$, and $R_{\rm max}$
are the same as those for the Bessel$\times$Gauss function.

\item[Gauss:]
a pure Gaussian
\begin{equation}
  c(r) = \left\{
    \begin{array}{ll}
      \exp \left[ -\left( \frac{r}{a} \right)^2 \right] &
      (r \le R_{\rm max}) \\
      0 & (\mbox{otherwise})
    \end{array} .
    \right.
\end{equation}
The default parameters are $a=1$, $R_{\rm max}=3$.

\item[Pillbox:]
a cell-averaging function
\begin{equation}
  c(x,y) = \left\{
    \begin{array}{ll}
      1 & (|x| \le 0.5, |y| \le 0.5) \\
      0 & (\mbox{otherwise})
    \end{array}
    \right.
\end{equation}
where $x$ and $y$ are the distance between
the data point and the grid point
(the unit is the grid spacing)
in the Cartesian coordinates along the map grid.

\item[Spheroidal:]
the spheroidal function
\begin{equation}
  c(x,y) = \left\{
    \begin{array}{ll}
      \multicolumn{2}{l}{|1-\eta_x^2|^\alpha
        \psi_{\alpha 0}(c,\eta_x) \times}\\
      \multicolumn{2}{c}{|1-\eta_y^2|^\alpha
        \psi_{\alpha 0}(c,\eta_y)} \\
        & ~~~(|\eta_x| \le 1, |\eta_y| \le 1) \\
      0 & ~~~(\mbox{otherwise})
    \end{array}
    \right.
\end{equation}
where $\eta_x=2x/m$, $\eta_y=2y/m$.
The parameters $m(=4, 5, 6, 7, 8)$ and
$\alpha(=0.0, 0.5, 1.0, 1.5, 2.0)$ define the shape of the function.
See \citet{schwab1984} for details.
The defaults are $m=6$ and $\alpha=1.0$.

\end{description}

The first four GCFs (Bessel$\times$Gauss, Sinc$\times$Gauss, Gauss,
and Pillbox) are taken from \citet{mangum2007}.
Given the cut-off radius $R_{\rm max}=3$,
the default parameters $a=1.55$, $b=2.52$ for the
Sinc$\times$Gauss function
are derived so that the energy concentration ratio ${\cal R}$
becomes the maximum \citep{schwab1984}.
Using the same parameters, ${\cal R}$ does not become
the very maximum for the Bessel$\times$Gauss function
(the maximum is found at $a\simeq 1.4$, $b\simeq 2.7$),
but is still high enough.

The spheroidal function is taken from \citet{schwab1984},
and is often used to regrid the interferometric visibility
($u$-$v$) data.
The function form is derived so that ${\cal R}$ becomes
as high as possible.

\begin{figure}
  \begin{center}
    \FigureFile(80mm,52mm){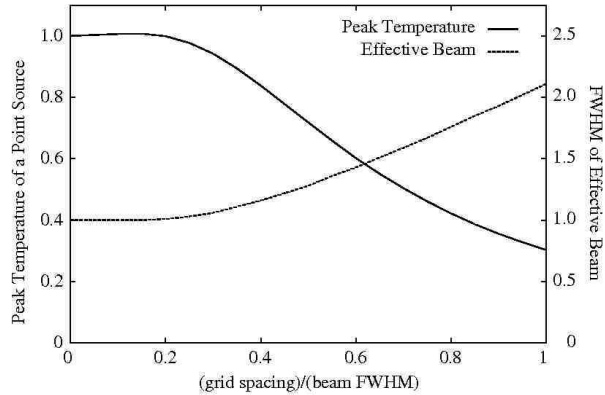}
  \end{center}
  \caption{Dependence of peak temperature of a point source
    ({\it left-hand axis\/}) and FWHM of effective beam
    ({\it right-hand axis\/}) on the grid spacing.
    Beam smearing effect is neglected.
    The vertical axes are normalized by the original (no convolution)
    values.
    The telescope beam is assumed to be a pure Gaussian,
    and the Bessel$\times$Gauss convolution with
    default parameters $a=1.55$, $b=2.52$, and $R_{\rm max}=3$
    is used.
  }\label{fig:convbeam}
\end{figure}

As a result of convolution, spatial resolution of the map
becomes lower than the telescope beam.
We estimate the influence of the beam broadening by
calculating the response to a point source.
Fig.\ \ref{fig:convbeam} shows the peak temperature
and full-width at half-maximum (FWHM) of the source
(i.e., effective beam)
on the obtained map, as functions of the grid spacing $d$.
If $d$ is too small, effective integration time for a grid
($t_{\rm cell}^{\rm ON}$) becomes small,
which leads to a large noise level.
In this case the map is too much oversampled,
since the spatial resolution is limited by the telescope beam.
On the other hand, if $d$ is too large, the effective beam
broadens to $\simeq 2d$.

\section{Verification}\label{sec:verification}

\subsection{Performance of the Antennas}\label{subsec:antperformance}

In \S \ref{sec:overview}, we introduced two parameters
$t_{\rm app}$ and $t_{\rm tran}$, which are duration of
``approach-run'' and ``transit-run'', respectively.
They depend on the driving speed and performance of the antenna.
In this subsection we describe the measurements to determine
$t_{\rm app}$ and $t_{\rm tran}$.
The amount of pointing jitter (${\rm PROG}-{\rm REAL}$) is
also measured.

\subsubsection{NRO 45-m}

\begin{figure}
  \begin{center}
    \FigureFile(80mm,58mm){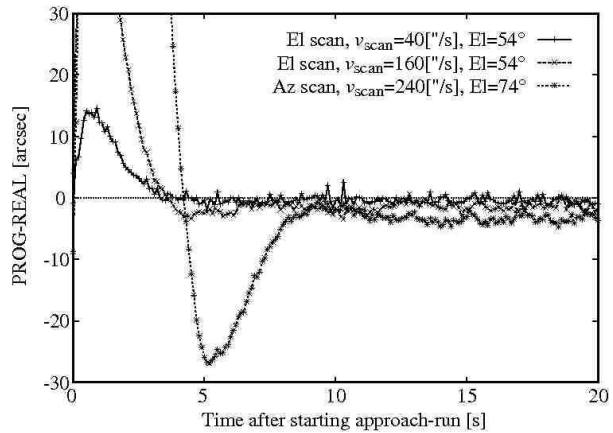}
  \end{center}
  \caption{The ${\rm PROG}-{\rm REAL}$ difference of
    the 45-m telescope
    along the scan direction on the sky is shown
    against the time elapsed after starting the approach-run.
    The measurements were done at every 0.1 second.
  }\label{fig:tapp45}
\end{figure}

First, we made scans along both Azimuth (Az) and Elevation (El)
at various scan speed $v_{\rm scan}$, in order to determine
$t_{\rm app}$.
The driving speed of the antenna
is $v_{\rm scan}/\cos({\rm El})$ for Az scans,
and $v_{\rm scan}$ for El scans.
Fig.\ \ref{fig:tapp45} shows some examples.
The ${\rm PROG}-{\rm REAL}$ difference on the sky is plotted
against the time elapsed after starting the approach-run
for three cases
(a) El scan with $v_{\rm scan}=40\;[^{\prime\prime}/{\rm s}]$
at ${\rm El}=\timeform{54D}$;
(b) El scan with $v_{\rm scan}=160\;[^{\prime\prime}/{\rm s}]$
at ${\rm El}=\timeform{54D}$; and
(c) Az scan with $v_{\rm scan}=240\;[^{\prime\prime}/{\rm s}]$
at ${\rm El}=\timeform{74D}$
(the driving speed is $870^{\prime\prime}/{\rm s}$ along Az).
The case (a) represents the typical scan at 115 GHz,
while the case (c) corresponds to the maximum scan speed
at 22 GHz.
The antenna runs stably after the initial delay and
(for large driving speed) an overshoot.
The time spent until the stable run begins is adopted
to be $t_{\rm app}$.
Determined $t_{\rm app}$ for the three cases are, respectively,
5, 7, and 11 seconds.

\begin{figure}
  \begin{center}
    \FigureFile(80mm,57mm){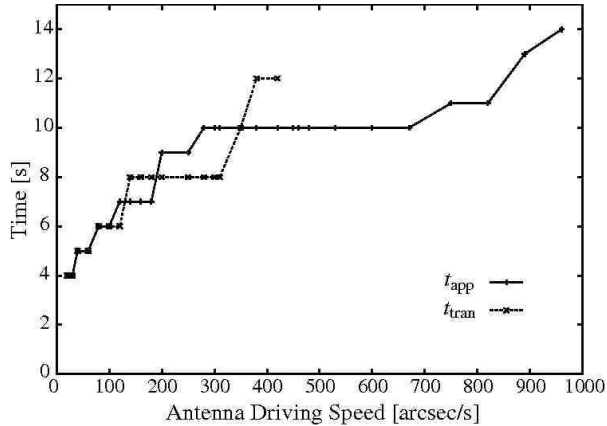}
  \end{center}
  \caption{Derived $t_{\rm app}$ and $t_{\rm tran}$ for
    the 45-m telescope.
    The horizontal axis shows the driving speed,
    $v_{\rm scan}/\cos({\rm El})$ for Az scans
    and $v_{\rm scan}$ for El scans.
  }\label{fig:tappttran45}
\end{figure}

Secondly, $t_{\rm tran}$ is determined in a similar way as
$t_{\rm app}$: back-and-forth scans along Az and El are made
using appropriate $t_{\rm app}$.
Various $t_{\rm tran}$ are tried.
The optimal $t_{\rm tran}$ is defined as the shortest one
for which the antenna starts to move stably after
the next approach-run.
Fig.\ \ref{fig:tappttran45} shows derived $t_{\rm app}$ and
$t_{\rm tran}$.
The both of them monotonically increase with the driving speed.
A slight jump of $t_{\rm app}$ is found at the driving speed of
$\simeq 200\;[^{\prime\prime}/{\rm s}]$.
It is due to ``slow-start slow-stop'' control of the telescope,
which is implemented not to give sudden and large acceleration
to the antenna.

It is found that the ${\rm PROG}-{\rm REAL}$ value does not
converge into 0 when the driving speed is large
(see Fig.\ \ref{fig:tapp45}).
Instead, ${\rm PROG}-{\rm REAL}$ offset becomes almost constant,
which amounts to
$|{\rm PROG}-{\rm REAL}| \simeq 0.016\;[{\rm s}]\times v_{\rm scan}$.
The offset does not affect the observations, since it is about $1/6$
of the sampling separation along the scan for $t_{\rm dump}$ of
0.1 second.
Excepting the offset, the jitter of the antenna pointing is
within a few arcseconds.
The error of the sampling separation is determined by the differential
of the jitter between the neighboring sample, which is almost
within \timeform{1"}--\timeform{2"}.

\subsubsection{ASTE 10-m}

\begin{figure}
  \begin{center}
    \FigureFile(80mm,58mm){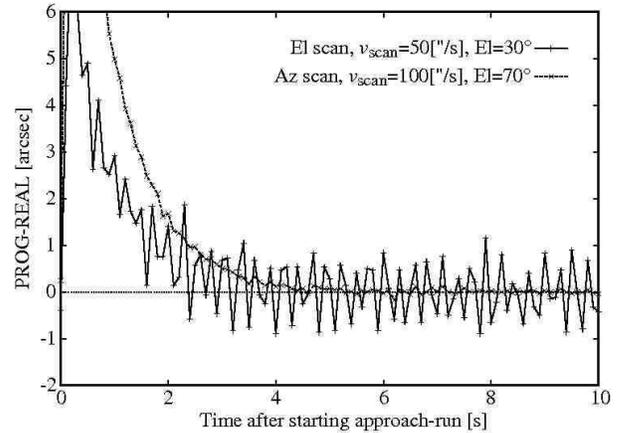}
  \end{center}
  \caption{Same as Fig.\ \ref{fig:tapp45}, but for the 10-m telescope.
  }\label{fig:tappaste}
\end{figure}

For the ASTE 10-m telescope, $t_{\rm app}$, $t_{\rm tran}$, and
the pointing jitter are measured in a similar way as the 45-m.
The $t_{\rm app}$ measurements for the cases
(a) El scan with $v_{\rm scan}=50\;[^{\prime\prime}/{\rm s}]$
at ${\rm El}=\timeform{30D}$ and
(b) Az scan with $v_{\rm scan}=100\;[^{\prime\prime}/{\rm s}]$
at ${\rm El}=\timeform{70D}$
(the driving speed is $290^{\prime\prime}/{\rm s}$ along Az)
are shown in Fig.\ \ref{fig:tappaste}.
The case (b) corresponds to the maximum scan speed at 350 GHz.
The ${\rm PROG}-{\rm REAL}$ values converge into $\simeq 0$
in a few seconds, with jitters of $\lesssim \pm \timeform{1"}$.
From this result, we adopted 4 seconds as $t_{\rm app}$.
The $t_{\rm tran}$ is also measured: for practical observing
parameters, 2 seconds are enough.
If receivers for higher frequency is installed in the future,
the antenna performance should be measured again.

\subsection{Comparison with PSW Map}\label{subsec:comparepsw}

In order to confirm the validity and efficiency of OTF observations,
the same field has been observed with both OTF and PSW,
and the resultant maps are compared.
Here we describe ASTE CO $J=3\mbox{--}2$ (345.8 GHz) test observations
toward a field centered at
$(l,b)=(\timeform{37D45'},\timeform{-0D12'})$.

The observations were made in 2005 August and September,
as a part of a CO $J=3-2$ Galactic plane survey
(Sawada et al., in preparation).
The frontend was a cooled SIS mixer receiver, SC345.
The system noise temperature was typically 200--300 K in
double sideband (DSB) during the observations.
The HPBW of the telescope at 350 GHz was measured to be \timeform{22"}.
The main beam efficiency was $\approx 0.6$.
The backend was a 1024-channel MAC,
which covers an instantaneous bandwidth of 512 MHz
($440\;{\rm km\; s^{-1}}$) with a spectral resolution of
1.0 MHz ($0.87\;{\rm km\; s^{-1}}$).
The pointing of the telescope was calibrated by tracking a compact
CO source W Aql in every 1 or 2 hours,
and was within the accuracy of \timeform{5"}.
We performed intensity calibration by observing a standard source,
M17 SW, in every 2 hours.
The reproducibility of $T_{\rm A}^*$ was 5\% ($1\sigma$).
By comparing the observed spectra with those measured with the
Caltech Submillimeter Observatory (CSO) 10.4-m telescope
with a single-sideband (SSB) filter \citep{wang1994},
we obtained scaling factors to convert ASTE
$T_{\rm A}^*({\rm DSB})$ into CSO $T_{\rm A}^*({\rm SSB})$.
Hereafter $T_{\rm A}^*$ is shown in SSB scale.

The OTF observations were carried out with parameters
$v_{\rm scan}=\timeform{50"}/{\rm s}$ and
${\mit\Delta}l=\timeform{8"}$.
Two longitudinal scans and two latitudinal scans
(in the Galactic coordinates) were made.
We made a map, whose grid spacing is
$\timeform{8"}\times\timeform{8"}\times 1\;{\rm km\;s^{-1}}$,
using the Bessel$\times$Gauss convolution;
and then the map was resampled onto a
$\timeform{10"}\times\timeform{10"}$ grid to match the PSW data.
A small portion of the region was observed with PSW:
$9\times 9$ points separated by \timeform{10"}.
For each point 10-second integration was made twice.
We have convolved the PSW data to make a map having the same 
resolution (\timeform{25"}) and grid spacing as the OTF map.
Since the convolution is highly incomplete at the outermost
grid points, we use the inner $7\times 7$ pixels for comparison.

\begin{figure}
  \begin{center}
    \FigureFile(80mm,58mm){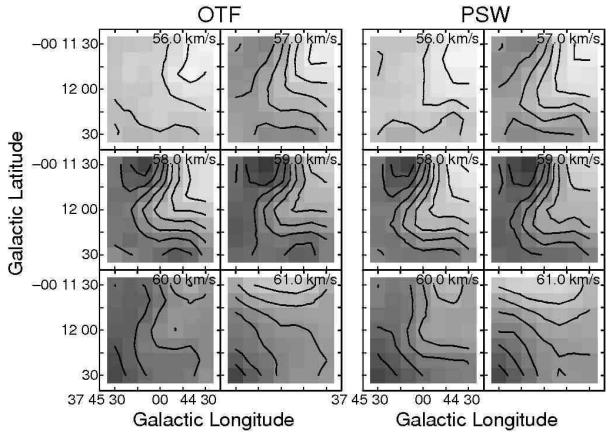}
  \end{center}
  \caption{Velocity channel maps obtained from OTF ({\it left\/}) and
    PSW ({\it right\/}) observations.
    The velocity range of
    $55.5\le v_{\rm LSR}\le 61.5\;{\rm km\;s^{-1}}$
    is sliced into ${\mit\Delta}v=1\;{\rm km\;s^{-1}}$ channels.
    Contours are $T_{\rm A}^*=1$ to 9 K at an interval of 1 K.
  }\label{fig:OTFvsPSWch}
\end{figure}

\begin{figure}
  \begin{center}
    \FigureFile(60mm,60mm){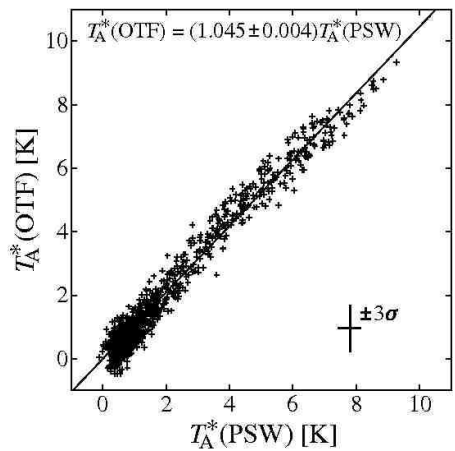}
  \end{center}
  \caption{A correlation plot between PSW and OTF $T_{\rm A}^*$
    in the velocity range $50\le v_{\rm LSR} \le 70\;{\rm km\;s^{-1}}$.
    The error of $\pm 3\sigma$ is shown in bottom right.
    Solid line shows the best fit
    $T_{\rm A}^*({\rm OTF}) = (1.045\pm 0.004) T_{\rm A}^*({\rm PSW})$.
  }\label{fig:PSWvsOTF}
\end{figure}

The $1\sigma$ noise levels are 0.25 K (OTF) and 0.12 K (PSW).
The obtained OTF noise level agrees with the one derived using
the equations in \S \ref{sec:parameters}, 0.26 K.
It is proven that the system has achieved
expected observing efficiency.
Velocity channel maps of OTF and PSW are shown in
Fig.\ \ref{fig:OTFvsPSWch}.
The OTF map is consistent with the PSW one.
Fig.\ \ref{fig:PSWvsOTF} shows a pixel-to-pixel correlation plot
between them.
A least-square fit gives the correlation
$T_{\rm A}^*({\rm OTF}) = (1.045\pm 0.004) T_{\rm A}^*({\rm PSW})$.
It is confirmed that the OTF map agrees with the PSW map
within the accuracy of relative intensity calibration, 5\%
(reproducibility of the intensity of the standard source).

\subsection{FSW Observations}

We have implemented FSW observing mode in OTF as described in
\S \ref{subsubsec:freqsw}.
As a test of FSW-OTF observations, IRAS 04369+2539 (IC 2087)
in Taurus molecular cloud was observed.
High-velocity wing (outflow) emission was found around the source
in the $^{12}{\rm CO}\;J=1\mbox{--}0$ line \citep{heyer1987}.
Mapping wing emission in widespread molecular cloud
is a presumable science case for FSW observations.
Stable spectral baseline is required.

Observations were carried out with the 45-m telescope and BEARS.
The HPBW is \timeform{15"}, and
the system noise temperature was typically 350 K (DSB).
The MAC was used in high-resolution mode, i.e.,
having 32 MHz ($83\;{\rm km\;s^{-1}}$) instantaneous bandwidth
and 63 kHz ($0.16\;{\rm km\;s^{-1}}$) resolution.
The OFFs are taken at starting point of approach-runs
in order to reduce dead time to slew the antenna.
The frequency throw of the LO, ${\mit\Delta}\nu_{\rm LO}$,
was set to 12 MHz (corresponding to $31\;{\rm km\;s^{-1}}$).
The $T_{\rm A}^*({\rm DSB})$ was converted into
$T_{\rm A}^*({\rm SSB})$ by comparing spectra of
a standard source measured with BEARS and an SSB receiver, S100.
Linear baselines were subtracted, and a
\timeform{7".5}-grid map was made
using the Bessel$\times$Gauss convolution.

\begin{figure}
  \begin{center}
    \FigureFile(70mm,139mm){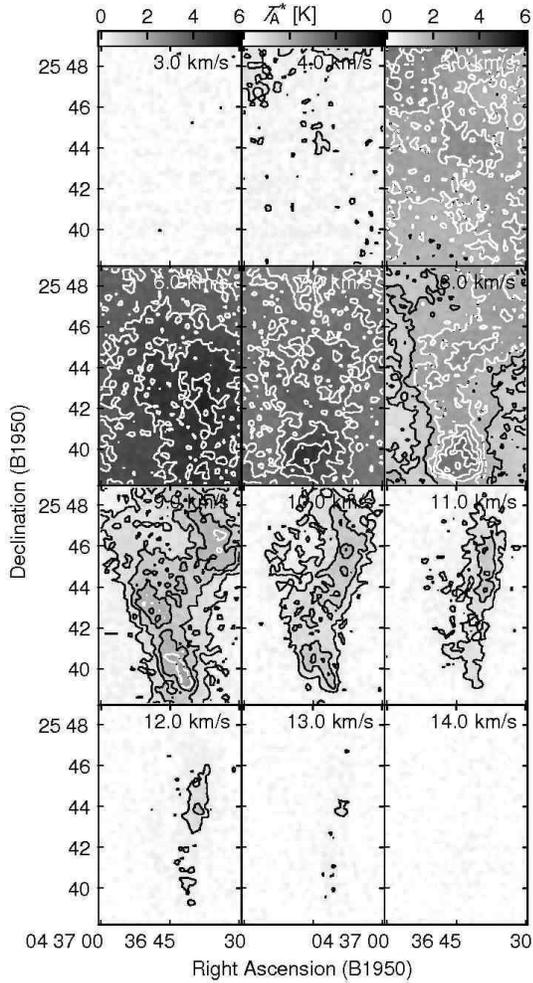}
  \end{center}
  \caption{Velocity channel maps of IC 2087
    $^{12}{\rm CO}\; J=1\mbox{--}0$
    emission at an interval of $1\;{\rm km\;s^{-1}}$.
    Contour levels are $T_{\rm A}^*=0.5$, 1.0, 1.5, ..., 5 K.
  }\label{fig:i2087ch}
\end{figure}

\begin{figure}
  \begin{center}
    \FigureFile(80mm,58mm){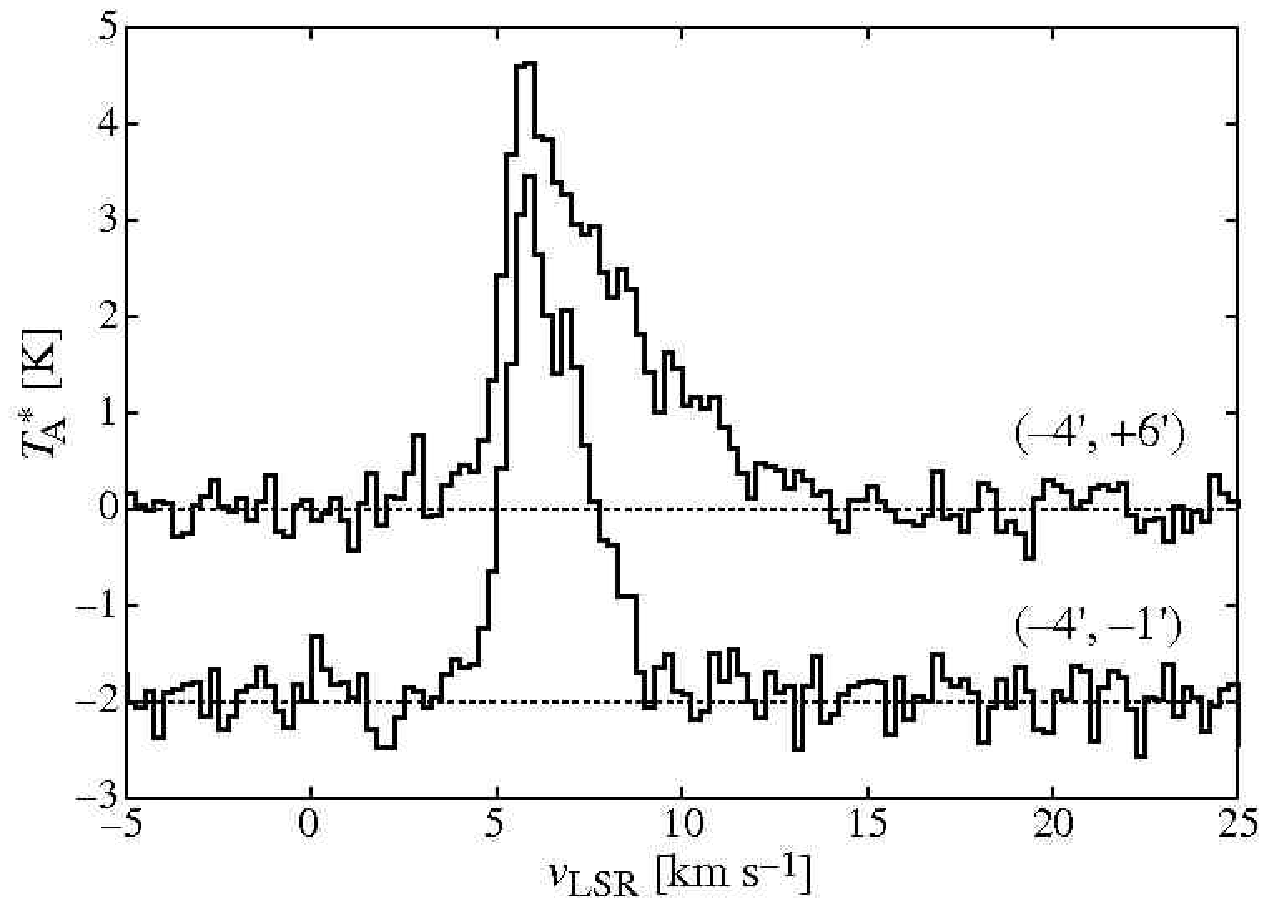}
  \end{center}
  \caption{The CO $J=1\mbox{--}0$ line profiles at two positions
    $({\mit\Delta}\alpha,{\mit\Delta}\delta)=(
    \timeform{-4'},\timeform{+6'})$ [{\it top\/}] and
    $(\timeform{-4'},\timeform{-1'})$
    [{\it bottom\/}, $-2\;{\rm K}$ offset].
    The positions are relative to
    $(\alpha,\delta)_{\rm 1950} =
    (\timeform{4h36m54s.6},\timeform{25D39'17"})$
  }\label{fig:i2087prof}
\end{figure}

Fig.\ \ref{fig:i2087ch} shows a set of velocity channel maps.
Redshifted wing emission toward north-south direction
is successfully detected beyond the ambient cloud velocity,
$v_{\rm LSR}\simeq 6\;{\rm km\;s^{-1}}$.
Fig.\ \ref{fig:i2087prof} shows line profiles at two positions
$({\mit\Delta}\alpha,{\mit\Delta}\delta)=
(\timeform{-4'},\timeform{+6'})$ and
$(\timeform{-4'},\timeform{-1'})$.
At $(\timeform{-4'},\timeform{+6'})$, wing emission extending to
$\simeq 14\;{\rm km\;s^{-1}}$ is seen, which is consistent with
Heyer's results.
On the other hand the profile at $(\timeform{-4'},\timeform{-1'})$,
which sharply truncates at $\simeq 9\;{\rm km\;s^{-1}}$,
demonstrates a straight baseline in a velocity range of
$40\;{\rm km\;s^{-1}}$.

\subsection{Application to Weak/Broad Line Objects}

In general, OTF is effective in particular when widely distributed
and intense line is mapped, since a beam runs across a map grid
within very short duration,
typically $(\mbox{a few}\mbox{--}10)\times 0.1$ seconds,
without any overhead to point discrete positions.
However, application to relatively small-field
($\sim$ a few arcminutes square), weak- and/or broad-line
($\sim$ several 10 mK, several 100 ${\rm km\;s^{-1}}$) sources
(e.g., external galaxies) has been successful.
For example, Hirota et al.\ (2008, to be submitted in this volume)
observed a galaxy IC 342 using the NRO 45-m and BEARS, and
achieved 8 mK rms in $T_{\rm A}^*$ at a velocity resolution of
$5\;{\rm km\; s^{-1}}$.

\section{Summary}

We have made spectral line OTF observations available
at NRO 45-m and ASTE 10-m telescopes.
Digital autocorrelation spectrometers can be operated
in OTF mode (the data sampling interval is as fast as 0.1 second)
with heterodyne receivers mounted on the telescopes,
including the 25-beam array receiver, BEARS.
Improvements of the software and instruments to enable
fast and synchronized controls (e.g., antenna driving,
data acquisition, Doppler tracking, frequency switching)
were described.
Sensitivity of the obtained map was expressed using observing
parameters, and we showed how to determine and optimize the parameters.
Performance of the antennas was measured and
was proven to be high enough for
up to 115 GHz (45-m) or 350 GHz (10-m) observations.
The OTF system has improved observing efficiency
by a factor of $\simeq 2$ compared with PSW:
its mapping capability opens a prospect in various
fields of study.

\bigskip

We thank NRO staffs and the ASTE team members for their support.
In particular, we acknowledge Satoshi Hongo, Aya Higuchi, Rie Miura,
Nobuyuki Yamaguchi, and Kunihiko Tanaka for their contribution
to development and evaluation.



\end{document}